\documentclass{article}
\usepackage{amsmath,graphicx}
\usepackage{spconf}
\usepackage{pdfpages}
\usepackage{enumitem}
\usepackage{multirow}
\setlist{nosep}

\title{Evidence of Vocal Tract Articulation in Self-Supervised Learning of Speech}
\name{Cheol Jun Cho$^{1}$\qquad Peter Wu$^{1}$\qquad Abdelrahman Mohamed$^{2}$\qquad  Gopala K. Anumanchipalli$^1$} 

\address{$^1$UC Berkeley, EECS, CA\\$^2$Meta AI}
\begin{document}

\maketitle
\begin{abstract}
Recent self-supervised learning (SSL) models  have proven to learn rich representations of speech, which can readily be utilized by diverse downstream tasks. To understand such utilities, various analyses have been done for speech SSL models to reveal which and how information is encoded in the learned representations. Although the scope of previous analyses is extensive in acoustic, phonetic, and semantic perspectives, the physical grounding by speech production has not yet received full attention. To bridge this gap, we conduct a comprehensive analysis to link speech representations to articulatory trajectories measured by electromagnetic articulography (EMA). Our analysis is based on a linear probing approach where we measure articulatory score as an average correlation of linear mapping to EMA. We analyze a set of SSL models selected from the leaderboard of the SUPERB benchmark \cite{yang2021superb} and perform further layer-wise analyses on two most successful models, Wav2Vec 2.0 \cite{Baevski2020} and HuBERT \cite{Hsu2021}. Surprisingly, representations from the recent speech SSL models are highly correlated with EMA traces (best: \(r = 0.81\)), and only 5 minutes are sufficient to train a linear model with high performance (\(r= 0.77\)). Our findings suggest that SSL models learn to align closely with continuous articulations, and provide a novel insight into speech SSL.

\end{abstract}
\begin{keywords}
Speech, Self-supervised learning, Electromagnetic articulography (EMA), Speech representation, Probing analysis, Acoustic-to-articulatory inversion
\end{keywords}
\section{Introduction}
\label{sec:intro}

Self-supervised learning (SSL) has been suggested as a pre-training method to learn representations without requiring a huge amount of labeled data. Recently proposed SSL models for speech provide rich representations which can be readily utilized for a broad range of spoken language tasks. When fine-tuned to downstream tasks, the SSL-based models are able to surpass supervised-only models \cite{Mohamed2022,yang2021superb}. Understanding how SSL models work is crucial to explain such success and to improve speech SSL. Previous studies have revealed acoustic, phonetic, and semantic representations encoded in representations of speech SSL, using diverse analytic tools including (non-) linear probing, mutual information, or canonical correlation analysis \cite{Baevski2021, Hsu2021, Pasad2021,  ma2021probing,  Shah2021}. Although those analyses have provided insights of information process in speech SSL models, the models still largely remain black box. Since the advent of data-driven deep learning strategies
for spoken language engineering, they have grown increasingly disconnected from human mechanisms and insights from speech production. 

\begin{figure}[!t]
  \includegraphics[width=8.6cm,keepaspectratio]{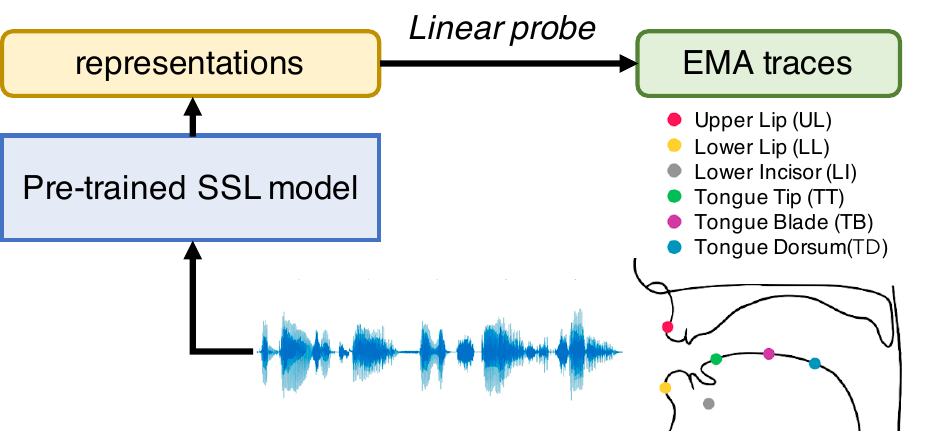}
  \caption{General framework of our analysis approach: linear probing of representations from pre-trained SSL models on EMA}
  \label{framework}
\end{figure}

To bridge this gap, we answer the question of how well the SSL representations are aligned with principles of speech production. We conduct a linear probing analysis of SSL representations against electromagnetic articulography (EMA) (Fig. \ref{framework}). EMA measures real-time continuous displacements of 6 articulators (Fig. \ref{framework}), which are timely synced with the produced speech \cite{mochatimit,richmond2011announcing}. As EMA tracks the actual physical dynamics of vocal tract, the modality is suitable for investigating physical grounding of speech representation. Moreover, a large portion of speech features are naturally subsumed by articulatory trajectories and a full speech can be reconstructed from EMA \cite{wu2022deep}. Underlying neurobiological process of speech production can also be explained by vocal tract articulation \cite{chartier2018encoding}. Therefore, we target EMA traces as principled and grounded reference for probing representations of SSL models. We firstly introduce articulatory score as an average correlation of linear prediction and EMA. Then, we conduct a comprehensive analysis by controlling model space, data size, and layers for probing. Our work bridges the gap between the current leading representation learning schemes and classical
speech production research.

Our major findings are:
\begin{itemize}
\item SSL representations are highly correlated with articulations: a simple linear regression is able to yield high correlation above 0.8.
\item Ranking by articulatory scores is coherent with the SUPERB leaderboard \cite{yang2021superb}, suggesting EMA prediction as a valid benchmark task.
\item Articulatory encoding of SSL representations is extremely robust that only 5 minutes of EMA data are sufficient to obtain high performance.
\item Layer-wise analysis reveals a shared pattern in Transformer based models except for last few layers, which is consistent with previous phoneme encoding analyses \cite{Baevski2020,Hsu2021,Pasad2021}. This trend is intact after fine-tuning for automatic speech recognition (ASR) task.

\end{itemize}

\section{Prior works}
\label{sec:priorwork}

The speech SSL benchmarking provides a general sense of representational power of SSL models, based on the downstream task performance \cite{yang2021superb}. Several studies have probed the learned representations themselves \cite{Pasad2021, Shah2021, Baevski2021, Hsu2021, ma2021probing}. \cite{Baevski2021} and \cite{Hsu2021} analyzes layer-wise phoneme encoding of their developed unsupervised models.  
\cite{Pasad2021} characterizes across layer patterns of acoustic, phonetic, and semantic encoding in speech SSL model. \cite{Shah2021} conducted non-linear probing analyses using extensive and fine-grained labels of speech features. Articulations are also included in the scope of \cite{Shah2021, ma2021probing}, but those are hand labeled pronunciations which are discrete and lack context dependent information such as coarticulation. Up to our knowledge, none of the prior works directly deals with continuous articulations represented as EMA. Unlike the previous probing analyses focus on a limited set of models, this study covers a broad range of speech models from hand-engineered features to the current state-of-the-art SSL models.

\section{Method}
\label{sec:method}

\subsection{Articulatory Datasets}
\label{ssec:artdata}

The articulatory dataset is composed of two EMA datasets, MNGU0 \cite{richmond2011announcing} and MOCHA-TIMIT \cite{mochatimit}. Both datasets provide 12 EMA traces (X, Y midsagittal coordinates for each of 6 articulators) of spoken utterances. We use 1,189 utterances by a single speaker (S1) from MNGU0 and 470 utterances by each of 7 speakers (S2-8) from MOCHA-TIMIT. The total duration is up to 75 minutes for MNGU0 and 28 minutes in average for MOCHA-TIMIT. 100 utterances of MNGU0 speaker and 50 utterances of each MOCHA-TIMIT speaker were held out as test sets. The data are normalized by channel and down-sampled to 50 Hz to match the sampling rate of SSL models.


\subsection{SSL models}
\label{ssec:sslmodel}

\begin{figure}[!t]
  \includegraphics[width=8.6cm,keepaspectratio]{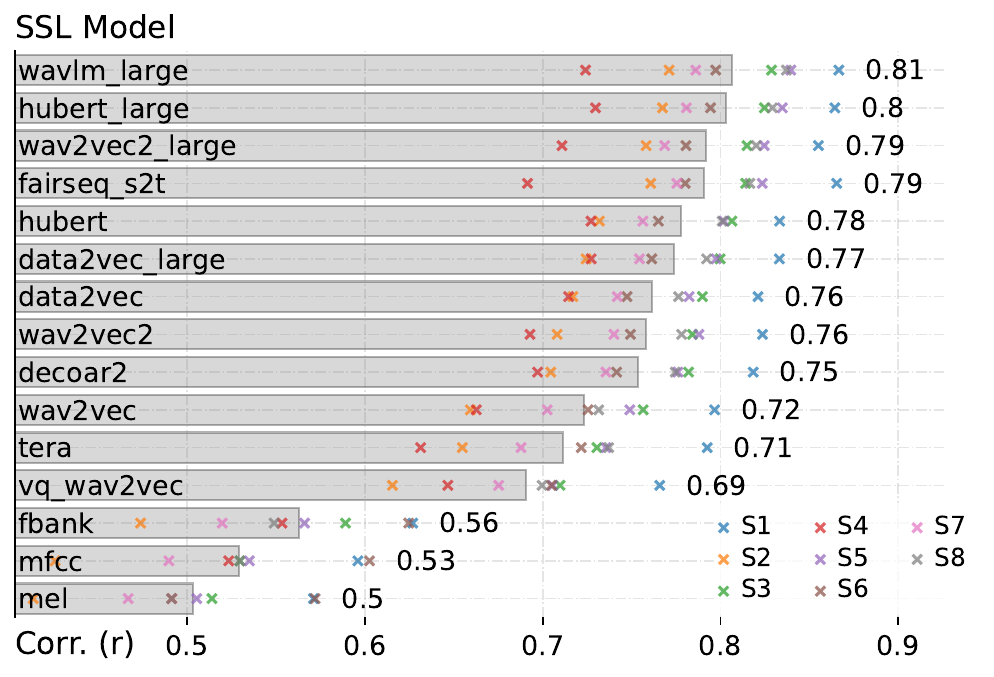}
  \caption{Articulatory score for each model. Each individual speaker's scores are marked with color. The annotated numbers indicate the scores averaged across speakers.}
  \label{modelcomp}
\end{figure}

We evaluate a pool of self-supervised speech representations by retrieving official models using S3PRL \cite{yang2021superb} and Huggingface.\footnote{\label{s3prl}https://github.com/s3prl/s3prl, \label{huggingface}  https://huggingface.co/} To give a fair comparison, we attempted to constrain the scope of analysis to the models pre-trained on LibriSpeech 960 hours corpus \cite{panayotov2015librispeech} or LibriLight 60K hours corpus \cite{kahn2020libri}, and we exclude extensions or variants. Although WavLM \cite{chen2022wavlm} is an extension of HuBERT \cite{Hsu2021}, the large version of the model is included as a reference of the best score we can get from SSL models. As a supervised counterpart, the encoder of Fairseq S2T \cite{wang2020fairseqs2t} is added, which was trained for ASR using LibriSpeech. Finally, some handcrafted acoustic features are included as the baselines. The final list includes: (\emph{contrastive model}) Wav2Vec \cite{schneider2019wav2vec}, VQ-Wav2Vec \cite{baevski2019vq} and  Wav2Vec 2.0 \cite{Baevski2020}, (\emph{predictive model}) HuBERT, Data2Vec \cite{Baevski2022} and WavLM, (\emph{generative model}) TERA \cite{liu2021tera} and DeCoAR 2.0 \cite{decoar2}, (\emph{supervised model}) Fairseq S2T, and (\emph{hand crafted}) filter bank, mel-spectrogram and MFCC (Fig. \ref{modelcomp}). If applicable, both base and large versions are included.

The large models were trained on LibriLight 60K hours except for WavLM which was trained on 57\% more data from other corpus in addition to LibriLight. All the other models used LibriSpeech for training.
By feeding speech audio of the EMA dataset, we extract representations from each layer of the models. For some models, features are down-/up-sampled to 50 Hz, so all representations are paired with synchronized EMA targets with the consistent sampling rate.

\subsection{Articulatory score}
\label{ssec:score}

Articulatory score is defined as an average correlation of linear probing. For each representational space from Sec. \ref{ssec:sslmodel}, a linear model is trained to predict each of 12 EMA traces, frame by frame.\footnote{By solving ordinary least squares regression without regularization.} The correlation is measured by Pearson's correlation coefficient (r) on held-out test set. Finally, articulatory score is obtained by averaging over 12 EMA channels and 8 speakers, if no articulator or speaker is specified.

We train separate models for each speaker since the individual variability in EMA is not only affected by speaker identity but also, or more, affected by the sensor configuration. 

\section{Results}
\label{sec:result}

\begin{figure}[!t]
  \includegraphics[width=8.6cm,keepaspectratio]{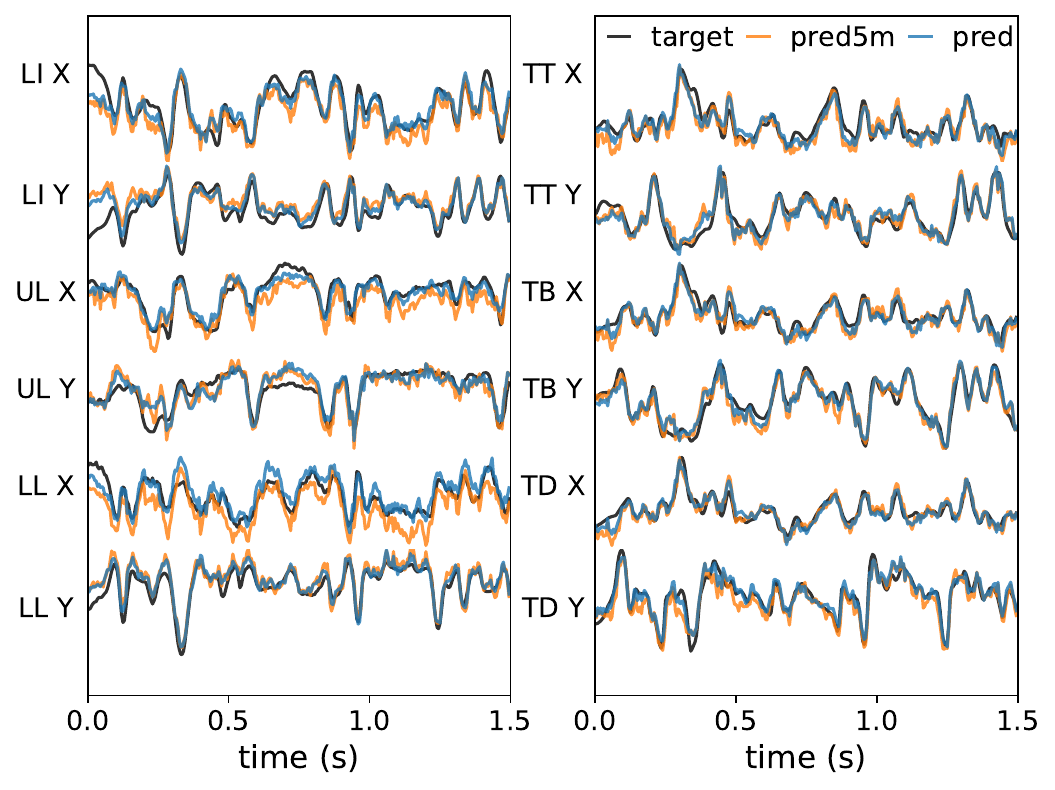}
  \caption{Example from test set. Separate models are trained on full training set (\emph{pred}) and 5 minute subset (\emph{pred5m}).}
  \label{prediction}
\end{figure}

\subsection{Comparison across SSL models}
\label{ssec:modelcomp}

To compare representational power of SSL models for articulations, we assess articulatory scores of SSL models. Highest scores are chosen among the multiple layers, and reported on Fig. \ref{modelcomp}. The overall pattern is consistent with SUPERB \cite{yang2021superb}, where Transformer based models trained with contrastive or predictive objectives (e.g., Wav2Vec 2.0, HuBERT, WavLM) are generally better than the other.
The ranks of the models mostly match those from the SUPERB leaderboard with high rank correlation (Kendall tau=0.82, Spearman=0.93). 
One unexpected deviation is Data2Vec Large, which shows lower score than the score of HuBERT Base. 

Overall scores are surprisingly high that even with base model such as HuBERT Base, we can get 0.78 correlation and for the best model, WavLM Large, the average correlation reaches 0.81, which are far beyond the handcrafted baselines (Fig. \ref{modelcomp}). Increasing model capacity can help build better articulatory representations, as the large models produce higher score than base versions. The supervised model, Fairseq S2T, has higher score than SSL models with comparable capacity, suggesting that ASR task is relevant for learning articulatory representations.

There is also a notable tendency across speakers that some speakers (e.g., S1, S5 and S8) show consistently high performance but some show low performance (e.g., S4, S2). This is due to noise ceiling of EMA data, where some speakers' data underwent unstable measurement.

\begin{figure}[!t]
  \includegraphics[width=8.6cm,keepaspectratio]{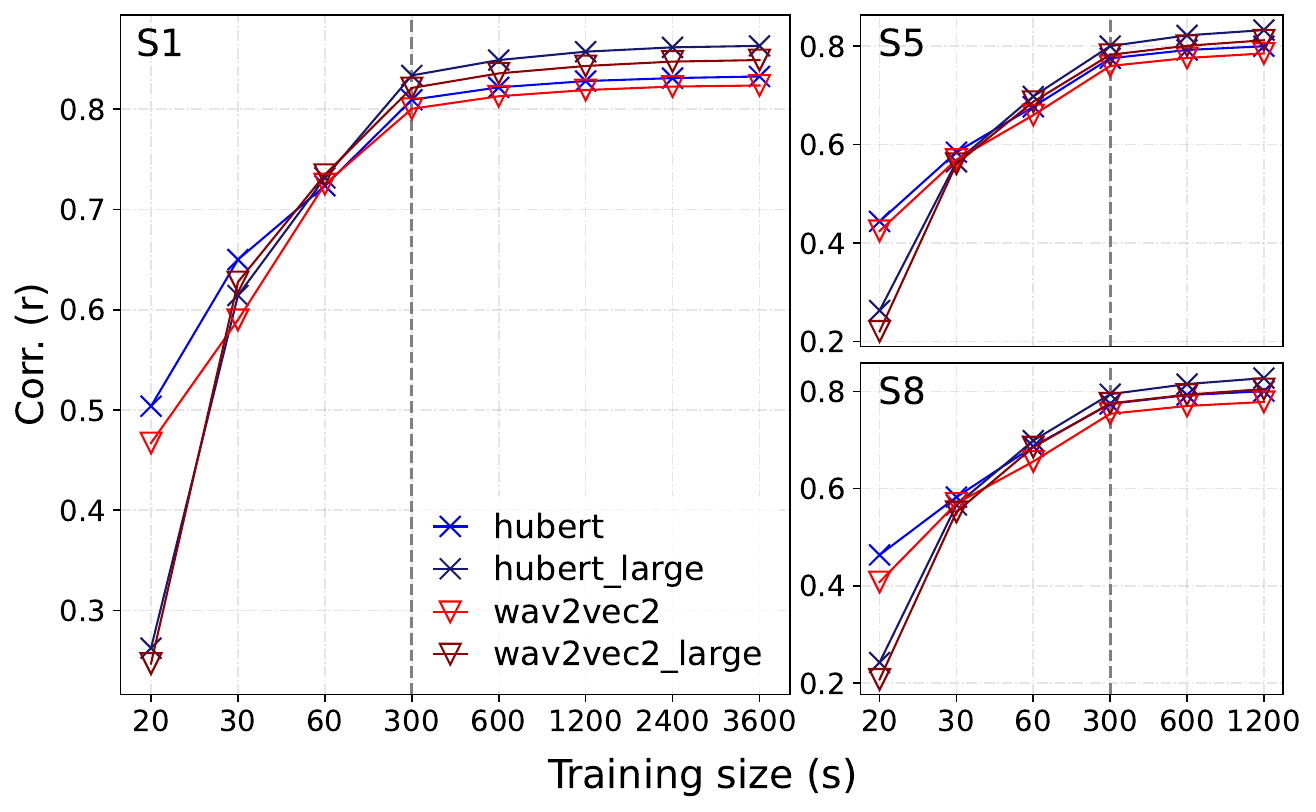}
  \caption{Articulatory scores by different amount of training data for three speakers: S1 (\emph{left}), S5 (\emph{top right}) and S8 (\emph{bottom right})}
  \label{trainsize}
\end{figure}

\subsection{Effects of training data size for EMA probing}
\label{ssec:sampleeff}
\begin{figure}[!t]
  \includegraphics[width=8.6cm,keepaspectratio]{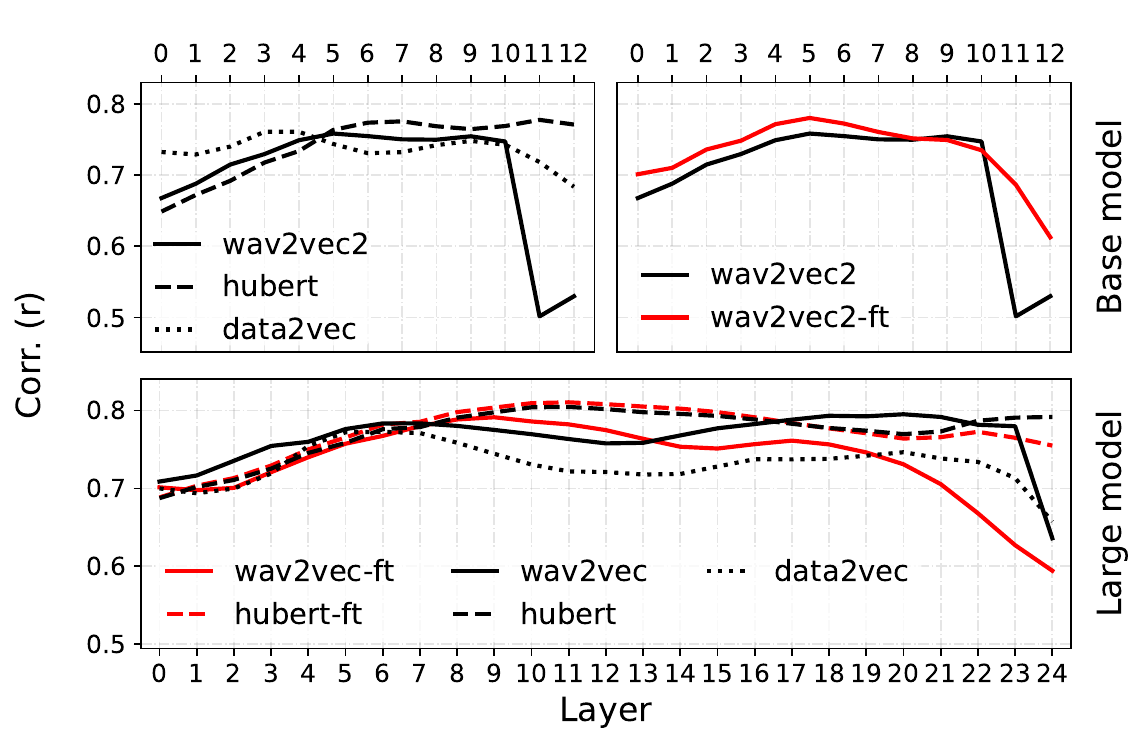}
  \caption{Layer-wise articulatory scores of base models (\emph{top left}), vs. fine-tuned model of Wav2Vec 2.0 (\emph{top right}) and any applicable large models (\emph{bottom}). Layer-0 is the output of CNN encoder. The red line, \emph{model}-ft, indicates the fine-tuned version of \emph{model}.}
  \label{layer}
\end{figure}

To further test how robustly representations are aligned with articulations, we control the size of training data for training linear models. The experiments were done using Wav2Vec 2.0 and HuBERT on three best speakers, S1, S5 and S8. The training size is one of [20s, 30s, 1m, 5m, 10m, 20m] and additional [40m, 60m] for S1. As shown in Fig. \ref{trainsize}, the scores increase dramatically until 5 minutes of training data and then start to saturate in all cases. This suggests that only 5 minutes of data are sufficient to build a high-performing linear mapping model. To verify this, we conducted the experiment with all 8 speakers using HuBERT Large layer-11 which shows the highest score among 24 layers (Fig. \ref{layer} \emph{bottom}).  The performance of different size of training data is reported on Table \ref{trainsizetable}. Strikingly, for all speakers, the extremely low-resource setting (5 min) shows sufficiently high-performance. This is also true when we qualitatively check the prediction (Fig. \ref{prediction}). This finding suggests that representations from the state-of-the-art SSL models efficiently encode a well generalizable representation of articulation.

\begin{table}[!t]
\centering
\begin{tabular}{|c|ccc|}
\hline
\multirow{2}{*}{Speaker} & \multicolumn{3}{c|}{Training size} \\ \cline{2-4} 
                         & 5 min         & 10 min       & All       \\ \hline
S1            & 0.83       & 0.85      & 0.87      \\
S2            & 0.74       & 0.76      & 0.77      \\
S3            & 0.79       & 0.81      & 0.82      \\
S4            & 0.67       & 0.70      & 0.73      \\
S5            & 0.80       & 0.82      & 0.83      \\
S6            & 0.76       & 0.78      & 0.79      \\
S7            & 0.74       & 0.76      & 0.78      \\
S8            & 0.80       & 0.82      & 0.83      \\ \hline
Average                  & 0.77       & 0.79      & 0.80      \\ \hline
\end{tabular}
\caption{Articulatory scores using HuBERT Large layer-11 trained on 5m, 10m and all training data.}
\label{trainsizetable}
\end{table}

\subsection{Encoding patterns across layer}
\label{ssec:layer}

 To investigate how information evolves through the layers, we evaluate articulatory score for each layer (Fig. \ref{layer}) from Wav2Vec 2.0, HuBERT and Data2Vec. These models were selected since they share model architecture with a convolutional waveform encoder followed by Transformer encoder, but these are trained with different learning objectives. 
 
 For all models, there are two local peaks in scores through the layers: the first one around the middle and the second one a few layers before the last layer (Fig. \ref{layer} \emph{top left}). This pattern is also found in large models and remained intact even after fine-tuning models for ASR on LibriSpeech 960 hours (Fig. \ref{layer} \emph{bottom and top right plots}). However, scores differ dramatically by models near the last layers where the effects of learning objective are strong. This is also supported by the observation that the later layers show larger changes than the earlier layers when fine-tuned. There are slight gains in performance after fine-tuning for all three models: Wav2Vec 2.0 Base 0.76 \(\rightarrow\) 0.78, Large 0.79 \(\rightarrow\) 0.80  and HuBERT Large 0.80 \(\rightarrow\) 0.81. Combined with the finding from Sec. \ref{ssec:modelcomp}, training for ASR task is contributing to learning articulatory representations. 
 
 Compared to previous probing analyses, the layer-wise patterns in Wav2Vec 2.0 and HuBERT are similar to those of the phoneme encoding analyses by \cite{Baevski2020, Hsu2021, Pasad2021}, which is natural since phonemes can be characterized by articulations.

\begin{figure}[!t]
  \includegraphics[width=8.6cm,keepaspectratio]{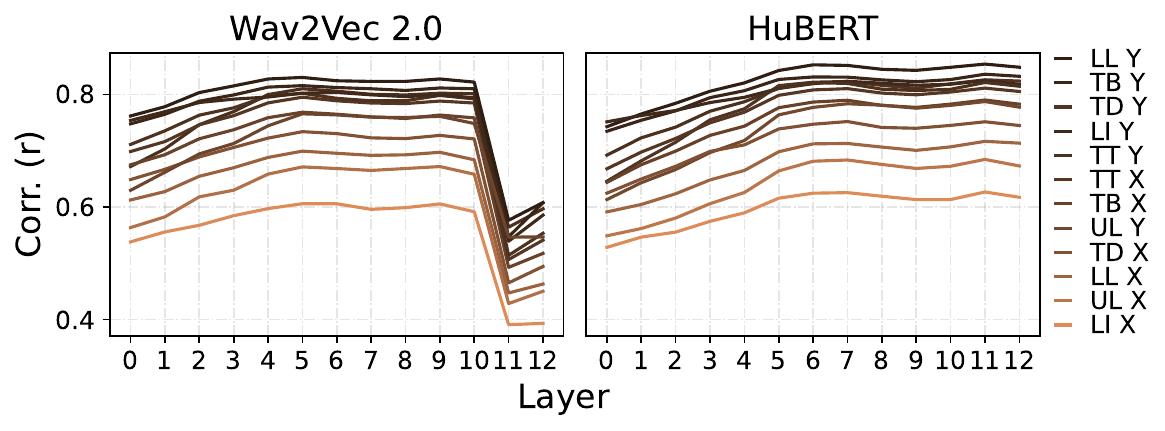}
  \caption{Layer-wise scores for each articulator from Wav2Vec 2.0  (\emph{left}) and HuBERT (\emph{right}) Each articulator is colored by average correlation.}
  \label{artlayer}
\end{figure}

Layer-wise scores for each articulator are plotted on Fig. \ref{artlayer}. The same across layer trend is consistently found in individual articulator for both models. There exists a variability among EMA channels: Y-coordinates of articulators are well-decoded while X-coordinates of lower incisor, lips, and tongue dorsum show relatively low scores. The articulators with low scores are indeed known to be less salient than other articulators in vocal tract dynamics.

\subsection{Speaker independent articulatory representations}
\label{ssec:speakerindep}

Although it is challenging to study speaker independent feature of EMA due to variable sensor configuration, we repeated experiments in two speaker independent schemes: 1) sharing models across speakers with the same train-test split, and 2) leave-one-speaker out cross-validation. In both settings, we were able to replicate the patterns reported in Sec. \ref{ssec:modelcomp} and Sec. \ref{ssec:layer}, with some degradation in scores.  
Since speaker identity gets marginalized in deep layers of SSL models \cite{chen2022speakerverif}, these consistent findings suggest that there exist speaker independent articulatory representations in SSL representations.

\section{Conclusion}
\label{sec:conclusion}

We probe recent SSL models with EMA traces and find that the learned representations have highly correlated and well-generalizable articulatory representations. Our findings suggest that the recent SSL models are well-grounded by the physical process of speech production. This motivates an interesting future direction of developing a more efficient model by directly grounding the model with articulations.

\section{Acknowledgements}

This research is supported by the following grants to PI Anumanchipalli --- NSF award 2106928, BAIR Commons-Meta AI, Rose Hills Foundation and Noyce Foundation.

\bibliographystyle{IEEEbib}
\fontsize{10}{11}\selectfont
\bibliography{refs}

\begin{thebibliography}{10}

\bibitem{yang2021superb}
Shu wen Yang, Po-Han Chi, Yung-Sung Chuang, et~al.,
\newblock ``{SUPERB: Speech Processing Universal PERformance Benchmark},''
\newblock in {\em Proc. Interspeech 2021}, 2021, pp. 1194--1198.

\bibitem{Baevski2020}
Alexei Baevski, Henry Zhou, Abdelrahman Mohamed, and Michael Auli,
\newblock ``wav2vec 2.0: A framework for self-supervised learning of speech
  representations,''
\newblock {\em Advances in Neural Information Processing Systems}, vol.
  2020-Decem, pp. 1--19, 2020.

\bibitem{Hsu2021}
Wei-Ning Hsu, Benjamin Bolte, Yao-Hung~Hubert Tsai, et~al.,
\newblock ``Hubert: Self-supervised speech representation learning by masked
  prediction of hidden units,''
\newblock {\em IEEE/ACM Transactions on Audio, Speech, and Language
  Processing}, vol. 29, pp. 3451--3460, 2021.

\bibitem{Mohamed2022}
Abdelrahman Mohamed, Hung-yi Lee, Lasse Borgholt, et~al.,
\newblock ``Self-supervised speech representation learning: A review,''
\newblock {\em arXiv preprint arXiv:2205.10643}, 2022.

\bibitem{Baevski2021}
Alexei Baevski, Wei-Ning Hsu, Alexis Conneau, and Michael Auli,
\newblock ``Unsupervised speech recognition,''
\newblock {\em Advances in Neural Information Processing Systems}, vol. 34, pp.
  27826--27839, 2021.

\bibitem{Pasad2021}
Ankita Pasad, Ju-Chieh Chou, and Karen Livescu,
\newblock ``Layer-wise analysis of a self-supervised speech representation
  model,''
\newblock in {\em 2021 IEEE Automatic Speech Recognition and Understanding
  Workshop (ASRU)}, 2021, pp. 914--921.

\bibitem{ma2021probing}
Danni Ma, Neville Ryant, and Mark Liberman,
\newblock ``Probing acoustic representations for phonetic properties,''
\newblock in {\em ICASSP 2021-2021 IEEE International Conference on Acoustics,
  Speech and Signal Processing (ICASSP)}. IEEE, 2021, pp. 311--315.

\bibitem{Shah2021}
Jui Shah, Yaman~Kumar Singla, Changyou Chen, and Rajiv~Ratn Shah,
\newblock ``What all do audio transformer models hear? probing acoustic
  representations for language delivery and its structure,''
\newblock {\em arXiv preprint arXiv:2101.00387}, 2021.

\bibitem{mochatimit}
Alan Wrench,
\newblock ``Mocha: multichannel articulatory database,''
  http://www.cstr.ed.ac.uk/research/project/\\artic/mocha.html, 1999.

\bibitem{richmond2011announcing}
Korin Richmond, Phil Hoole, and Simon King,
\newblock ``Announcing the electromagnetic articulography (day 1) subset of the
  mngu0 articulatory corpus,''
\newblock in {\em Twelfth Annual Conference of the International Speech
  Communication Association}, 2011.

\bibitem{wu2022deep}
Peter Wu, Shinji Watanabe, Louis Goldstein, Alan~W Black, and Gopala~K
  Anumanchipalli,
\newblock ``Deep speech synthesis from articulatory representations,''
\newblock {\em arXiv preprint arXiv:2209.06337}, 2022.

\bibitem{chartier2018encoding}
Josh Chartier, Gopala~K Anumanchipalli, Keith Johnson, and Edward~F Chang,
\newblock ``Encoding of articulatory kinematic trajectories in human speech
  sensorimotor cortex,''
\newblock {\em Neuron}, vol. 98, no. 5, pp. 1042--1054, 2018.

\bibitem{panayotov2015librispeech}
Vassil Panayotov, Guoguo Chen, Daniel Povey, and Sanjeev Khudanpur,
\newblock ``Librispeech: an asr corpus based on public domain audio books,''
\newblock in {\em 2015 IEEE international conference on acoustics, speech and
  signal processing (ICASSP)}. IEEE, 2015, pp. 5206--5210.

\bibitem{kahn2020libri}
Jacob Kahn, Morgane Rivi{\`e}re, Weiyi Zheng, Evgeny Kharitonov, Qiantong Xu,
  et~al.,
\newblock ``Libri-light: A benchmark for asr with limited or no supervision,''
\newblock in {\em ICASSP 2020-2020 IEEE International Conference on Acoustics,
  Speech and Signal Processing (ICASSP)}. IEEE, 2020, pp. 7669--7673.

\bibitem{chen2022wavlm}
Sanyuan Chen, Chengyi Wang, Zhengyang Chen, et~al.,
\newblock ``Wavlm: Large-scale self-supervised pre-training for full stack
  speech processing,''
\newblock {\em IEEE Journal of Selected Topics in Signal Processing}, 2022.

\bibitem{wang2020fairseqs2t}
Changhan Wang, Yun Tang, Xutai Ma, et~al.,
\newblock ``fairseq s2t: Fast speech-to-text modeling with fairseq,''
\newblock in {\em Proceedings of the 2020 Conference of the Asian Chapter of
  the Association for Computational Linguistics (AACL): System Demonstrations},
  2020.

\bibitem{schneider2019wav2vec}
Steffen Schneider, Alexei Baevski, Ronan Collobert, and Michael Auli,
\newblock ``wav2vec: Unsupervised pre-training for speech recognition,''
\newblock {\em Proc. Interspeech 2019}, pp. 3465--3469, 2019.

\bibitem{baevski2019vq}
Alexei Baevski, Steffen Schneider, and Michael Auli,
\newblock ``vq-wav2vec: Self-supervised learning of discrete speech
  representations,''
\newblock in {\em International Conference on Learning Representations}, 2019.

\bibitem{Baevski2022}
Alexei Baevski, Wei-Ning Hsu, Qiantong Xu, et~al.,
\newblock ``Data2vec: A general framework for self-supervised learning in
  speech, vision and language,''
\newblock {\em arXiv preprint arXiv:2202.03555}, 2022.

\bibitem{liu2021tera}
Andy~T Liu, Shang-Wen Li, and Hung-yi Lee,
\newblock ``Tera: Self-supervised learning of transformer encoder
  representation for speech,''
\newblock {\em IEEE/ACM Transactions on Audio, Speech, and Language
  Processing}, vol. 29, pp. 2351--2366, 2021.

\bibitem{decoar2}
Shaoshi Ling and Yuzong Liu,
\newblock ``Decoar 2.0: Deep contextualized acoustic representations with
  vector quantization,''
\newblock {\em arXiv preprint arXiv:2012.06659}, 2020.

\bibitem{chen2022speakerverif}
Sanyuan Chen, Yu~Wu, Chengyi Wang, et~al.,
\newblock ``Why does self-supervised learning for speech recognition benefit
  speaker recognition?,''
\newblock {\em arXiv preprint arXiv:2204.12765}, 2022.

\end{thebibliography}
\label{sec:refs}
\end{document}